\documentstyle[12pt]{article}

\newcommand {\e} {\mbox{\rm e}}

\newcounter{eq}
\newcounter{sc}

\newcommand {\PR}   {Phys. Rev.}

\def\overleftrightarrow#1{\vbox{\ialign{##\crcr
 $\leftrightarrow$\crcr\noalign{\kern-1pt\nointerlineskip}
 $\hfil\displaystyle{#1}\hfil$\crcr}}}

\newlength{\minitwocolumn}
\begin{document}

%%%%%%%%%%%%%%%%%%%%%%%%%%%%%%%%%%%%%%%%%%%%%%%%%%%%%%%%%%%%%%%%%%
%%%%%%%%%%%%%%%%%%%%%%%% Title %%%%%%%%%%%%%%%%%%%%%%%%%%%%%%%%%%%
%%%%%%%%%%%%%%%%%%%%%%%%%%%%%%%%%%%%%%%%%%%%%%%%%%%%%%%%%%%%%%%%%%
\begin{flushright}
DPUR/TH/40\\
June, 2014\\
%hep-th/070****\\
\end{flushright}
\vspace{20pt}

%\magnification=\magstep1
\pagestyle{empty}
\baselineskip15pt
%\font\cmssB=cmss17
%\font\cmssS=cmss10

\begin{center}
{\large\bf Quadratic Chaotic Inflation from Higgs Inflation
\vskip 1mm }

\vspace{20mm}
Ichiro Oda \footnote{E-mail address:\ ioda@phys.u-ryukyu.ac.jp}
and Takahiko Tomoyose \footnote{E-mail address:\ k138335@eve.u-ryukyu.ac.jp}

\vspace{5mm}
           Department of Physics, Faculty of Science, University of the 
           Ryukyus,\\
           Nishihara, Okinawa 903-0213, Japan.\\

\end{center}

%\maketitle

\vspace{5mm}
\begin{abstract}
Stimulated with the recent discovery of B-mode by BICEP2, we discuss the relation 
between a Higgs inflation and a chaotic inflation with quadratic potential. Starting with 
a generalized Higgs inflation model, we derive a condition for obtaining the quadratic 
chaotic inflation. It is shown that the running of the Higgs self-coupling constant in the 
Jordan frame plays a decisive role when the generalized Higgs inflation model coincides 
with the Higgs inflation model in a small-field limit.  
\end{abstract}

\newpage
\pagestyle{plain}
\pagenumbering{arabic}
%\setcounter{page}{1}

%%%%%%%%%%%%%%%%%%%%%%%%%%%%%%%%%%%%%%%%%%%%%%%%%%%%%%%%%%%%%%%%%%
%%%%%%%%%%%%%%%%%%%%%%%% Article %%%%%%%%%%%%%%%%%%%%%%%%%%%%%%%%%
%%%%%%%%%%%%%%%%%%%%%%%%%%%%%%%%%%%%%%%%%%%%%%%%%%%%%%%%%%%%%%%%%%

\rm
%%%%%%%%%%%%%%%%%%%%%%%%%%%%%%%%%%%%%%%%%%%%%%%%%%%%%%%%%%%%%%%%%%%%%
%%%%%%%%%%%%%%%%%%%%%%%%%%%%%%   SEC  1    %%%%%%%%%%%%%%%%%%%%%%%%%%
%%%%%%%%%%%%%%%%%%%%%%%%%%%%%%%%%%%%%%%%%%%%%%%%%%%%%%%%%%%%%%%%%%%%%
\section{Introduction}

The BICEP2 experiment has recently announced a remarkable discovery of the primodial 
B-mode polarization in the cosmic microwave background (CMB) \cite{BICEP2}, thereby 
giving us a strong support for the inflation scenario \cite{Inflation}. According to the BICEP2
results \cite{BICEP2}, the tensor-to-scalar ratio is given by $r = 0.20^{+0.07}_{-0.05}$ 
(this value is decreased to $r = 0.16^{+0.06}_{-0.05}$ after subtracting the best available
estimate for foreground dust), which suggests that a large-field inflation has happened
at a very early stage of evolution of the universe. Then, inflation is sensitive to quantum
gravity and string theory, so for the first time we might have a chance of understanding
quantum gravitational effects experimentally by examining the CMB carefully. 

Of particular interest is that the BICEP2 results favor the simplest chaotic inflation \cite{Linde}
with quadratic potential 
%**   Quad-pot  %%%%%%%%%%%%%%%%%%%%%%%%%%%%%%%%%%%%%%%%%%%%%%%%%%%%%%%%%
\begin{eqnarray}
{\cal L} = - \frac{1}{2} ( \partial \phi )^2 - \frac{1}{2} m^2 \phi^2,
\label{Quad-pot}
\end{eqnarray}
%%%%%%%%%%%%%%%%%%%%%%%%%%%%%%%%%%%%%%%%%%%%%%%%%%%%%%%%%%%%%%%%%%%
where $\phi$ is an inflaton field and $m$ is its mass. An important issue here is
the identification of the inflaton field $\phi$.

As the most economical approach, it is tempting to identify the inflaton field $\phi$ with
the Higgs field in the standard model (SM), which has been found at the LHC recently
\cite{ATLAS, CMS}. Indeed, such an inflation model is nowadays called the Higgs inflation
which has been actively investigated so far \cite{Higgs-inflation}. A peculiar feature of
the Higgs inflation is the presence of a non-minimal coupling term of the inflaton to
gravity, by which the flatness of the potential at large values of the inflaton is ensured.
Once the non-minimal coupling constant is experimentally fixed by the amplitude of 
the scalar perturbations, the theory has a strong predictive power. For instance, 
this model predicts the spectral index $n_s  \approx 0.97$ and the tensor-to-scalar
ratio $r \approx 0.003$. It is worthwhile to recall that one of features in the Higgs inflation 
is a lowering of the tensor-to-scalar ratio. 
Actually, the prediction of the tensor-to-scalar ratio $r \approx 0.003$ by the Higgs inflation 
is obviously incompatible with the larger value  $r \approx 0.20$ by the BICEP2, which yielded 
a conjecture that the Higgs inflation might be dead \cite{Cook}. 

However, more recently, the paper entitled "Higgs inflation still alive" has appeared where 
a critical observation is that by tuning of the top quark mass one can make a saddle point 
in the Higgs potential \cite{Kawai1}. \footnote{See also a related work \cite{Bezrukov}.} 
Then, it is pointed out that the e-folding is earned by passing the saddle point, which leads to
the almost same  tensor-to-scalar ratio as that by BICEP2, and the observational density perturbation 
corresponds to the size of the Higgs field above the saddle point.

In this article, we wish to clarify the relation between the Higgs inflation and the chaotic inflation with
quadratic potential. The key point is that we must make use of not the constant but the running Higgs 
self-coupling constant with a minimum near the Planck mass scale  \cite{Kawai2}. 
Although there are two prescriptions identifying the renormalization scale using the Einstein frame 
and the Jordan one, leading to the same conclusion in Ref.  \cite{Kawai1}, our formalism
prefers the Jordan frame to the Einstein one.

This article is organized as follows: In the next section, we construct a generalized Higgs inflation in the 
Jordan frame which agrees with the conventional Higgs inflation at small values of the Higgs field.
In Section 3, we derive an equation between the generic functions by requiring the generalized Higgs inflation to
become equivalent to the quadratic chaotic inflation in the Einstein frame.
In Section 4, we consider the case of the original Higgs inflation and show that 
the running of the Higgs self-coupling constant in the Jordan frame plays a decisive role 
when the generalized Higgs inflation model coincides with the Higgs inflation model.
The final section is devoted to discussion.

%%%%%%%%%%%%%%%%%%%%%%%%%%%%%%%%%%%%%%%%%%%%%%%%%%%%%%%%%%%%%%%%%%%%%
%%%%%%%%%%%%%%%%%%%%%%%%%%%%%%   SEC  2    %%%%%%%%%%%%%%%%%%%%%%%%%%
%%%%%%%%%%%%%%%%%%%%%%%%%%%%%%%%%%%%%%%%%%%%%%%%%%%%%%%%%%%%%%%%%%%%%
\section{Generalized Higgs inflation}

Let us begin with the construction of the generalized Higgs inflation where compared to the
original Higgs inflation  \cite{Higgs-inflation} the factors in front of both the scalar curvature and the Higgs kinetic term
are extended to generic functions of the Higgs field. \footnote{A generalized Higgs inflation based on  
the Galieon field has been considered in \cite{Kamada}.}  A portion of the SM Lagrangian relevant to our argument
takes the following form:
%**   Lagr 1  %%%%%%%%%%%%%%%%%%%%%%%%%%%%%%%%%%%%%%%%%%%%%%%%%%%%%%%%%
\begin{eqnarray}
{\cal L} = \sqrt{-g} \left[ F(H) R - K(H) g^{\mu\nu} \partial_\mu H^\dagger \partial_\nu H
- U(H) \right],
\label{Lagr 1}
\end{eqnarray}
%%%%%%%%%%%%%%%%%%%%%%%%%%%%%%%%%%%%%%%%%%%%%%%%%%%%%%%%%%%%%%%%%%%
where $H$ is the Higgs doublet for which we take the unitary gauge $H = \frac{1}{\sqrt{2}} (0, h)^T$ throughout this article. 
Moreover, $U(H)$ is the conventional Higgs potential defined as
%**   Higgs-p  %%%%%%%%%%%%%%%%%%%%%%%%%%%%%%%%%%%%%%%%%%%%%%%%%%%%%%%%%
\begin{eqnarray}
U(H) = \lambda \left( H^\dagger H - \frac{v^2}{2} \right)^2 = \frac{\lambda}{4} \left( h^2 - v^2 \right)^2 \equiv U(h),
\label{Higgs-p}
\end{eqnarray}
%%%%%%%%%%%%%%%%%%%%%%%%%%%%%%%%%%%%%%%%%%%%%%%%%%%%%%%%%%%%%%%%%%%
with $\lambda$ and $v$ being the Higgs self-coupling constant and the vacuum expectation value of the Higgs
field, respectively.
 
In order to move from the Jordan frame to the Einstein frame, let us perform the "conformal"
transformation \footnote{Precisely speaking, this transformation is not the conformal one since
the Higgs field remains intact \cite{Oda}.} 
%**   Conf-transf %%%%%%%%%%%%%%%%%%%%%%%%%%%%%%%%%%%%%%%%%%%%%%%%%%%%%%%%%
\begin{eqnarray}
g_{\mu\nu} \rightarrow \tilde g_{\mu\nu} = \Omega^2(x) g_{\mu\nu},  \quad
g^{\mu\nu} \rightarrow \tilde g^{\mu\nu} = \Omega^{-2}(x) g^{\mu\nu}, 
\label{Conf-transf}
\end{eqnarray}
%%%%%%%%%%%%%%%%%%%%%%%%%%%%%%%%%%%%%%%%%%%%%%%%%%%%%%%%%%%%%%%%%%%
from which we can derive useful formulae:
%**   Formulae %%%%%%%%%%%%%%%%%%%%%%%%%%%%%%%%%%%%%%%%%%%%%%%%%%%%%%%%%
\begin{eqnarray}
\sqrt{-g} = \Omega^{-4} \sqrt{- \tilde g},  \quad
R = \Omega^2 ( \tilde R + 6 \tilde \Box f - 6 \tilde g^{\mu\nu} \partial_\mu f \partial_\nu f ), 
\label{Formulae}
\end{eqnarray}
%%%%%%%%%%%%%%%%%%%%%%%%%%%%%%%%%%%%%%%%%%%%%%%%%%%%%%%%%%%%%%%%%%%
where we have defined as $f = \log \Omega$ and $\tilde \Box f = \frac{1}{\sqrt{- \tilde g}} 
\partial_\mu (\sqrt{- \tilde g} \tilde g^{\mu\nu} \partial_\nu f) = \tilde g^{\mu\nu} 
\tilde \nabla_\mu \tilde \nabla_\nu f$.

With the choice of the scale factor $\Omega$ satisfying \footnote{Here $M_p = \sqrt{\frac{c \hbar}{8 \pi G}} 
= 2.44 \times 10^{18} GeV$ is the reduced Planck mass.}
%**   Choice %%%%%%%%%%%%%%%%%%%%%%%%%%%%%%%%%%%%%%%%%%%%%%%%%%%%%%%%%
\begin{eqnarray}
\Omega^2 = \frac{2 F(h)}{M_p^2},
\label{Choice}
\end{eqnarray}
%%%%%%%%%%%%%%%%%%%%%%%%%%%%%%%%%%%%%%%%%%%%%%%%%%%%%%%%%%%%%%%%%%%
it is easy to rewrite the Lagrangian (\ref{Lagr 1}) up to a surface term as 
%**   Lagr 2  %%%%%%%%%%%%%%%%%%%%%%%%%%%%%%%%%%%%%%%%%%%%%%%%%%%%%%%%%
\begin{eqnarray}
{\cal L} = \sqrt{- \tilde g} \left\{  \frac{1}{2} M_p^2 \tilde R - \frac{M_p^2}{4} 
\left[ \frac{K}{F} + 3 \left(\frac{F^\prime}{F} \right)^2 \right] 
\tilde g^{\mu\nu} \partial_\mu h \partial_\nu h - V(h) \right\},
\label{Lagr 2}
\end{eqnarray}
%%%%%%%%%%%%%%%%%%%%%%%%%%%%%%%%%%%%%%%%%%%%%%%%%%%%%%%%%%%%%%%%%%%
where prime means the derivative with respect to $h$, i.e., $F^\prime = \frac{d F}{d h}$ 
and the new potential in the Einstein frame is defined as 
%**   V  %%%%%%%%%%%%%%%%%%%%%%%%%%%%%%%%%%%%%%%%%%%%%%%%%%%%%%%%%
\begin{eqnarray}
V(h) = \frac{\lambda M_p^4}{16} 
\left(\frac{h^2 - v^2}{F(h)} \right)^2.
\label{V}
\end{eqnarray}
%%%%%%%%%%%%%%%%%%%%%%%%%%%%%%%%%%%%%%%%%%%%%%%%%%%%%%%%%%%%%%%%%%%
Provided that one introduces a new scalar field $\chi$ defined as 
%**  Chi1  %%%%%%%%%%%%%%%%%%%%%%%%%%%%%%%%%%%%%%%%%%%%%%%%%%%%%%%%%
\begin{eqnarray}
\frac{d \chi}{d h} = \frac{M_p}{\sqrt{2}} \frac{\sqrt{K(h) F(h) + 3 \left( F^\prime(h) \right)^2}}{F(h)},
\label{Chi1}
\end{eqnarray}
%%%%%%%%%%%%%%%%%%%%%%%%%%%%%%%%%%%%%%%%%%%%%%%%%%%%%%%%%%%%%%%%%%%
the Lagrangian (\ref{Lagr 2}) can be cast to
%**   Lagr 3  %%%%%%%%%%%%%%%%%%%%%%%%%%%%%%%%%%%%%%%%%%%%%%%%%%%%%%%%%
\begin{eqnarray}
{\cal L} = \sqrt{- \tilde g} \left[  \frac{1}{2} M_p^2 \tilde R 
- \frac{1}{2} \tilde g^{\mu\nu} \partial_\mu \chi \partial_\nu \chi - V(h(\chi)) \right].
\label{Lagr 3}
\end{eqnarray}
%%%%%%%%%%%%%%%%%%%%%%%%%%%%%%%%%%%%%%%%%%%%%%%%%%%%%%%%%%%%%%%%%%%
This is the Lagrangian of our generalized Higgs inflation in the Einstein frame. 
We will require this Lagrangian to become the quadratic chaotic inflation by 
restricting the form of the potential $V(h(\chi))$.

%%%%%%%%%%%%%%%%%%%%%%%%%%%%%%%%%%%%%%%%%%%%%%%%%%%%%%%%%%%%%%%%%%%%%
%%%%%%%%%%%%%%%%%%%%%%%%%%%%%%   SEC  3    %%%%%%%%%%%%%%%%%%%%%%%%%%
%%%%%%%%%%%%%%%%%%%%%%%%%%%%%%%%%%%%%%%%%%%%%%%%%%%%%%%%%%%%%%%%%%%%%
\section{Quadratic chaotic inflation}

Now we wish to have a chaotic inflation with quadratic potential from the above generalized
Higgs inflation (\ref{Lagr 3}). To this aim, the new
potential $V(h(\chi))$ must take the form
%**   Mass-term  %%%%%%%%%%%%%%%%%%%%%%%%%%%%%%%%%%%%%%%%%%%%%%%%%%%%%%%%%
\begin{eqnarray}
V(h(\chi)) = \frac{1}{2} m^2 \chi^2, 
\label{Mass-term}
\end{eqnarray}
%%%%%%%%%%%%%%%%%%%%%%%%%%%%%%%%%%%%%%%%%%%%%%%%%%%%%%%%%%%%%%%%%%%
where $m$ is the mass of the scalar field $\chi$. Together with Eqs.  (\ref{V}) and  (\ref{Mass-term}),
$\chi$ is expressed in terms of $h$ as 
%**   Chi2  %%%%%%%%%%%%%%%%%%%%%%%%%%%%%%%%%%%%%%%%%%%%%%%%%%%%%%%%%
\begin{eqnarray}
\chi = \frac{M_p^2}{2 m} \sqrt{\frac{\lambda}{2}} \frac{h^2 - v^2}{F(h)}.
\label{Chi2}
\end{eqnarray}
%%%%%%%%%%%%%%%%%%%%%%%%%%%%%%%%%%%%%%%%%%%%%%%%%%%%%%%%%%%%%%%%%%%

It is known that the Higgs self-coupling constant $\lambda$ has a peculiar behaviour as a function
of the renormalization scale $\mu$. For instance, around the Planck mass scale $M_p$, the running
coupling constant  $\lambda(\mu)$ has a minimum at $\mu_{min} \approx 10^{17-18} GeV$ depending
on the Higgs mass and can be approximated as 
%**   Running coupling1  %%%%%%%%%%%%%%%%%%%%%%%%%%%%%%%%%%%%%%%%%%%%%%%%%%%%%%%%%
\begin{eqnarray}
\lambda(\mu) = \lambda_{min} + \frac{b}{(16 \pi^2)^2} \left( log \frac{\mu}{\mu_{min}}  \right)^2,
\label{Running coupling1}
\end{eqnarray}
%%%%%%%%%%%%%%%%%%%%%%%%%%%%%%%%%%%%%%%%%%%%%%%%%%%%%%%%%%%%%%%%%%%
where $b  \approx 0.6$ \cite{Kawai1, Bezrukov, Salvio}. There are two prescriptions how to select
the renormalization scale $\mu$. One prescription is to choose the renormalization scale $\mu$ to be
the effective mass in the Einstein frame
%**   Prescrition1  %%%%%%%%%%%%%%%%%%%%%%%%%%%%%%%%%%%%%%%%%%%%%%%%%%%%%%%%%
\begin{eqnarray}
\mu = \frac{M_p}{\sqrt{2 F(h)}} h.
\label{Prescrition1}
\end{eqnarray}
%%%%%%%%%%%%%%%%%%%%%%%%%%%%%%%%%%%%%%%%%%%%%%%%%%%%%%%%%%%%%%%%%%%
The other  prescription is to choose it to be the effective mass in the Jordan frame
%**   Prescrition2  %%%%%%%%%%%%%%%%%%%%%%%%%%%%%%%%%%%%%%%%%%%%%%%%%%%%%%%%%
\begin{eqnarray}
\mu = h.
\label{Prescrition2}
\end{eqnarray}
%%%%%%%%%%%%%%%%%%%%%%%%%%%%%%%%%%%%%%%%%%%%%%%%%%%%%%%%%%%%%%%%%%% 
Even if both the prescriptions work well in Ref.  \cite{Kawai1}, we have to make use of the
latter prescription in this study as seen shortly. With this prescription, the running coupling
constant becomes a function of $h$ as 
%**   Running coupling2  %%%%%%%%%%%%%%%%%%%%%%%%%%%%%%%%%%%%%%%%%%%%%%%%%%%%%%%%%
\begin{eqnarray}
\lambda(\mu) = \lambda(h).
\label{Running coupling2}
\end{eqnarray}
%%%%%%%%%%%%%%%%%%%%%%%%%%%%%%%%%%%%%%%%%%%%%%%%%%%%%%%%%%%%%%%%%%%

Finally, let us note that the mathematical consistency requires the differentiation of $\chi$ in Eq.  (\ref{Chi2}) with
respect to $h$ to be equal to Eq.  (\ref{Chi1}). Using this consistency condition, we can express the function
$K(h)$ like  
%**  K  %%%%%%%%%%%%%%%%%%%%%%%%%%%%%%%%%%%%%%%%%%%%%%%%%%%%%%%%%
\begin{eqnarray}
K = \frac{\lambda M_p^2}{4 m^2} \frac{1}{F} \left[ \frac{1}{2} \frac{\lambda^\prime}{\lambda}  (h^2 - v^2) 
+ 2 h  - (h^2 - v^2)  \frac{F^\prime}{F} \right]^2 - 3 \frac{(F^\prime)^2}{F}.
\label{K}
\end{eqnarray}
%%%%%%%%%%%%%%%%%%%%%%%%%%%%%%%%%%%%%%%%%%%%%%%%%%%%%%%%%%%%%%%%%%%

%%%%%%%%%%%%%%%%%%%%%%%%%%%%%%%%%%%%%%%%%%%%%%%%%%%%%%%%%%%%%%%%%%%%%
%%%%%%%%%%%%%%%%%%%%%%%%%%%%%%   SEC  4    %%%%%%%%%%%%%%%%%%%%%%%%%%
%%%%%%%%%%%%%%%%%%%%%%%%%%%%%%%%%%%%%%%%%%%%%%%%%%%%%%%%%%%%%%%%%%%%%
\section{Higgs inflation}

Now let us consider the Higgs inflation in the Jordan frame by specifying the function $F(h)$ to be the following 
form \cite{Higgs-inflation} \footnote{Although it is possible to generalize the function $F(h)$ to a more 
general expression such as $F(h) = \frac{1}{2} M_p^2 \left(1 + \frac{\xi h^2}{n M_p^2}\right)^n$ which reduces to
(\ref{HI}) in the limit $\xi h^2 \ll M_p^2$, it turns out that we obtain a similar result to that
in case of (\ref{HI}).}
%**  HI  %%%%%%%%%%%%%%%%%%%%%%%%%%%%%%%%%%%%%%%%%%%%%%%%%%%%%%%%%
\begin{eqnarray}
F(h) = \frac{1}{2} (M_p^2 + \xi h^2). 
\label{HI}
\end{eqnarray}
%%%%%%%%%%%%%%%%%%%%%%%%%%%%%%%%%%%%%%%%%%%%%%%%%%%%%%%%%%%%%%%%%%%
With the choice (\ref{HI}), the function $K(h)$ in Eq. (\ref{K}) is given by
%**  K-HI-1  %%%%%%%%%%%%%%%%%%%%%%%%%%%%%%%%%%%%%%%%%%%%%%%%%%%%%%%%%
\begin{eqnarray}
K = \frac{\lambda M_p^2}{2 m^2 (M_p^2 + \xi h^2)} \left[ \frac{1}{2} \frac{\lambda^\prime}{\lambda}  (h^2 - v^2) 
+ \frac{2 (M_p^2 + \xi v^2)}{M_p^2 + \xi h^2} h \right]^2 - \frac{6 \xi^2 h^2}{M_p^2 + \xi h^2}.
\label{K-HI-1}
\end{eqnarray}
%%%%%%%%%%%%%%%%%%%%%%%%%%%%%%%%%%%%%%%%%%%%%%%%%%%%%%%%%%%%%%%%%%%
 
Since the running coupling constant $\lambda$ is a complicated and numerical function, we shall confine ourselves 
to the analysis of the function  $K(h)$ only in the limits of both $\xi h^2 \gg M_p^2$
and $\xi h^2 \ll M_p^2$. In the limit $\xi h^2 \gg M_p^2$, since $\lambda$ is approximated as in 
Eq. (\ref{Running coupling1}), $K$ is reduced to
%**  K-HI-2  %%%%%%%%%%%%%%%%%%%%%%%%%%%%%%%%%%%%%%%%%%%%%%%%%%%%%%%%%
\begin{eqnarray}
K \rightarrow k_1 \equiv \frac{b}{2 (16 \pi)^2 \xi} \left(\frac{M_p}{m}\right)^2 - 6 \xi.
\label{K-HI-2}
\end{eqnarray}
%%%%%%%%%%%%%%%%%%%%%%%%%%%%%%%%%%%%%%%%%%%%%%%%%%%%%%%%%%%%%%%%%%% 
The requirement that the Higgs particle is not a ghost but a normal particle imposes the constraint
that $k_1$ in Eq. (\ref{K-HI-2}) should be positive definite. Then, the non-minimal coupling
constant $\xi$ is bounded by
%**  Xi-1  %%%%%%%%%%%%%%%%%%%%%%%%%%%%%%%%%%%%%%%%%%%%%%%%%%%%%%%%%
\begin{eqnarray}
\xi < \frac{1}{32 \pi} \sqrt{\frac{b}{3}} \frac{M_p}{m} \approx 7 \times 10^2,
\label{Xi-1}
\end{eqnarray}
%%%%%%%%%%%%%%%%%%%%%%%%%%%%%%%%%%%%%%%%%%%%%%%%%%%%%%%%%%%%%%%%%%% 
where we have put $b = 0.6, \ m = 1.5 \times 10^{13} GeV, \ M_p = 2.44 \times 10^{18} GeV$.
This inequality certainly covers the value $\xi \approx 10$ in Ref. \cite{Kawai1}.
With the redefinitions $\sqrt{k_1} h = \bar h, \ \sqrt{k_1} v = \bar v, \ \frac{1}{k_1^2} \lambda 
= \bar \lambda$ and $\frac{1}{k_1} \xi = \bar \xi$, the starting Lagrangian (\ref{Lagr 1})
can be rewritten in the limit $\xi h^2 \gg M_p^2$ as
%**   Lagr HI-1  %%%%%%%%%%%%%%%%%%%%%%%%%%%%%%%%%%%%%%%%%%%%%%%%%%%%%%%%%
\begin{eqnarray}
{\cal L} \approx \sqrt{-g} \left[ \frac{1}{2} (M_p^2 + \bar \xi \bar h^2) R 
- \frac{1}{2} g^{\mu\nu} \partial_\mu \bar h \partial_\nu \bar h
- \frac{\bar \lambda}{4} ( \bar h^2 - \bar v^2)^2 \right],
\label{Lagr HI-1}
\end{eqnarray}
%%%%%%%%%%%%%%%%%%%%%%%%%%%%%%%%%%%%%%%%%%%%%%%%%%%%%%%%%%%%%%%%%%%
which is nothing but the Lagrangian of the Higgs inflation. It is of interest to note
that the Lagrangian of the generalized Higgs inflation takes the same form as that of the Higgs inflation 
even in the region of large field.

Next we turn our attention to the opposite limit $\xi h^2 \ll M_p^2$. In this limit, the Lagrangian (\ref{Lagr 1})
must become that of the Higgs inflation in order to ensure that our model in fact describes the Higgs inflation. 
The running coupling constant in this limit is approximated as 
%**   Running coupling3  %%%%%%%%%%%%%%%%%%%%%%%%%%%%%%%%%%%%%%%%%%%%%%%%%%%%%%%%%
\begin{eqnarray}
\lambda(h) = \lambda_0 \e^{- \alpha h},
\label{Running coupling3}
\end{eqnarray}
%%%%%%%%%%%%%%%%%%%%%%%%%%%%%%%%%%%%%%%%%%%%%%%%%%%%%%%%%%%%%%%%%%%
where $\lambda_0$ and $\alpha$ are positive constants. Then, $K$ reads
%**  K-HI-3  %%%%%%%%%%%%%%%%%%%%%%%%%%%%%%%%%%%%%%%%%%%%%%%%%%%%%%%%%
\begin{eqnarray}
K \rightarrow k_2 \equiv 2 \lambda_0 \left(\frac{v}{m}\right)^2,
\label{K-HI-3}
\end{eqnarray}
%%%%%%%%%%%%%%%%%%%%%%%%%%%%%%%%%%%%%%%%%%%%%%%%%%%%%%%%%%%%%%%%%%% 
which is positive definite as desired. \footnote{This result is robust against changes in the approximation of 
the running coupling constant. For instance, 
if we approximate the running coupling constant in this limit by using a quadratic function as
%**   Running coupling4  %%%%%%%%%%%%%%%%%%%%%%%%%%%%%%%%%%%%%%%%%%%%%%%%%%%%%%%%%
\begin{eqnarray}
\lambda(h) = \lambda_0 (h - h_0)^2,
\label{Running coupling4}
\end{eqnarray}
%%%%%%%%%%%%%%%%%%%%%%%%%%%%%%%%%%%%%%%%%%%%%%%%%%%%%%%%%%%%%%%%%%%
where $\lambda_0$ and $h_0$ are some positive constants (this equation holds for $h < h_0$),  $K$ is given by
%**  K-HI-4  %%%%%%%%%%%%%%%%%%%%%%%%%%%%%%%%%%%%%%%%%%%%%%%%%%%%%%%%%
\begin{eqnarray}
K \rightarrow k_2 \equiv 2 \lambda_0 \left[\frac{(v - h_0) v}{m}\right]^2,
\label{K-HI-4}
\end{eqnarray}
%%%%%%%%%%%%%%%%%%%%%%%%%%%%%%%%%%%%%%%%%%%%%%%%%%%%%%%%%%%%%%%%%%% 
which is positive definite as well.}  Thus, with the redefinitions 
$\sqrt{k_2} h = \hat h, \ \sqrt{k_2} v = \hat v, \ \frac{1}{k_2^2} \lambda 
= \hat \lambda$ and $\frac{1}{k_2} \xi = \hat \xi$, it is certainly true that in the limit $\xi h^2 \ll M_p^2$
the Lagrangian (\ref{Lagr 1}) can be cast to be the same form as that of the Higgs inflation
%**   Lagr HI-2  %%%%%%%%%%%%%%%%%%%%%%%%%%%%%%%%%%%%%%%%%%%%%%%%%%%%%%%%%
\begin{eqnarray}
{\cal L} \approx \sqrt{-g} \left[ \frac{1}{2} (M_p^2 + \hat \xi \hat h^2) R 
- \frac{1}{2} g^{\mu\nu} \partial_\mu \hat h \partial_\nu \hat h
- \frac{\hat \lambda}{4} ( \hat h^2 - \hat v^2)^2 \right].
\label{Lagr HI-2}
\end{eqnarray}
%%%%%%%%%%%%%%%%%%%%%%%%%%%%%%%%%%%%%%%%%%%%%%%%%%%%%%%%%%%%%%%%%%% 

%%%%%%%%%%%%%%%%%%%%%%%%%%%%%%%%%%%%%%%%%%%%%%%%%%%%%%%%%%%%%%%%%%%%%
%%%%%%%%%%%%%%%%%%%%%%%%%%%%%%   SEC  4    %%%%%%%%%%%%%%%%%%%%%%%%%%
%%%%%%%%%%%%%%%%%%%%%%%%%%%%%%%%%%%%%%%%%%%%%%%%%%%%%%%%%%%%%%%%%%%%%
\section{Discussion}

In this article, we have clarified the relation between the Higgs inflation and the quadratic chaotic inflation
when the running of the Higgs self-coupling constant is switched on. Via the construction of the generalized
Higgs inflation model, we have shown that the Higgs inflation in the Jordan frame can be described by quadratic chaotic
inflation in the Einstein frame in treating with both the running coupling constant and the kinetic term
in a proper manner.

Moreover, it is shown that within the present framework, the Lagrangian of the generalized Higgs inflation 
is reduced to that of the usual Higgs inflation in both large and small field limits. 
To do that, it is essential to choose the renormalization 
scale $\mu$ to be the Higgs field $h$, which is done only in the Jordan frame. In this sense, in the formalism at hand,
the prescription in the Jordan frame possesses a more preferred position than that in the Einstein frame.

From our work, it is found that the Higgs inflation with the running of the Higgs self-coupling constant
and  the generalized Higgs kinetic term in the Jordan frame can be precisely rewritten as a quadratic chaotic inflation
in the Einstein frame, which nicely explains the recent results by BICEP2 \cite{BICEP2}.  Usually, in the Higgs inflation
to account for the BICEP2 results \cite{Kawai1, Bezrukov}, even if the running of the Higgs self-coupling constant
is taken into consideration, the kinetic term of the Higgs field is kept to be canonical, so it is difficult to
rewrite the Lagrangian in the Jordan frame to that of the quadratic chaotic inflation in the Einstein frame.
Thus, our study might shed a light on the importance of the modification of the Higgs kinetic term
in regarding the Higgs field as the inflaton.  

As a future work, we wish to report the unitarity issue \cite{Unitarity} in the present formalism.

%%%%%%%%%%%%%%%%%%%%%%%%%%%%%%%%%%%%%%%%%%%%%%%%%%%%%%%%%%%%%%%%%%
%%%%%%%%%%%%%%%%%%%%%%%% Acknowledgements %%%%%%%%%%%%%%%%%%%%%%%%%%%%%
%%%%%%%%%%%%%%%%%%%%%%%%%%%%%%%%%%%%%%%%%%%%%%%%%%%%%%%%%%%%%%%%%%
\begin{flushleft}
{\bf Acknowledgements}
\end{flushleft}
This work is supported in part by the Grant-in-Aid for Scientific 
Research (C) Nos. 22540287 and 25400262 from the Japan Ministry of Education, Culture, 
Sports, Science and Technology.

%%%%%%%%%%%%%%%%%%%%%%%% reference %%%%%%%%%%%%%%%%%%%%%%%%%%%%%%%
%%%%%%%%%%%%%%%%%%%%%%%%%%%%%%%%%%%%%%%%%%%%%%%%%%%%%%%%%%%%%%%%%%

\end{document}